\documentclass[pra, twocolumn, floatfix]{revtex4-2}
\usepackage{graphicx}
\usepackage{epstopdf}
\usepackage{amsmath, amsfonts, amssymb, bm,color}

\def\ifa{Institute of Applied Physics, Moldova State University,
Academiei str. 5, MD-2028 Chi\c{s}in\u{a}u, Moldova}

\begin{document}
\title{Dense dipole-dipole-coupled two-level systems in a thermal bath}
\author{Mihai A. Macovei }
\email{mihai.macovei@ifa.usm.md}
\affiliation{\ifa}
\date{\today}
\begin{abstract}
The quantum dynamics of a dense and dipole-dipole coupled ensemble of two-level emitters interacting via their environmental 
thermostat is investigated. The static dipole-dipole interaction strengths are being considered strong enough but smaller than 
the transition frequency. Therefore, the established thermal equilibrium of ensemble's quantum dynamics is described with 
respect to the dipole-dipole coupling strengths. We have demonstrated the quantum nature of the spontaneously scattered 
light field in this process for weaker thermal baths as well as non-negligible dipole-dipole couplings compared to the emitter's 
transition frequency. Furthermore, the collectively emitted photon intensity suppresses or enhances depending on the 
environmental thermal baths intensities.
\end{abstract}
\maketitle

\section{Introduction}
The mutual interactions among multiple excited two-level emitters are being mediated by their environmental vacuum electromagnetic 
field reservoir. Depending on the inter-particle separations, these interactions diminish or can be enhanced, respectively 
\cite{dicke,gsag,supm1,puri,colar,colar1}. Certain of these emission-absorption collective features were demonstrated experimentally 
which resulted, for instance, in superradiance and subradiance \cite{supr,subr,supbr,suprk}, superabsorption \cite{qab,supab} as well 
as in various phase transitions phenomena \cite{supm2,pDM}, etc \cite{coll1,coll2,stls,coll3,susyl}. In a thermal electromagnetic field 
environmental reservoir, the quantum dynamics of a small two-level ensemble changes accordingly. Particularly, it was demonstrated 
that the entanglement between two initially independent qubits can be generated if the two qubits interact with a common heat bath 
in thermal equilibrium \cite{ent1,ent2,ent3,ent4}. Larger two-level ensembles, under the Dicke approximation \cite{dicke,gsag,supm1,puri} 
and in a common thermostat, obey the Bose-Einstein statistics and there are no critical steady-state behaviours in the thermodynamic 
limit which are typical to laser-driven atomic ensembles \cite{puri,supm2,pDM}. Moreover, quantum light features are proper only for 
few atoms in these setups \cite{hassan,mm}. Contrary, the electromagnetic fields emitted on the two transitions of a three-level 
Dicke-like ensemble, surrounded by the thermostat, are strongly correlated or anticorrelated and exhibit quantum light properties 
also for larger numbers of atoms. This is demonstrated via violation of the Cauchy-Schwarz inequality \cite{mek}.

The multi-qubit collective phenomena, mediated by environmental thermal baths, have attracted an additional interest recently 
due to feasible applications related to quantum thermodynamics \cite{qth1,qth2,qth3,qth4}. In this respect, the performance 
of quantum heat engines \cite{qhe1,qhe2,qhe3,qhe4,qhe5,qhe6}, quantum refrigerators \cite{qref1,qref2} or quantum 
batteries \cite{qb1,qb2,qb3} has been shown to improve considerably due to cooperativity among their constituents. 

Motivated by these advances regarding multi-qubit ensembles and their applications, here, we investigate the quantum dynamics 
of a small dipole-dipole interacting two-level ensemble in thermal equilibrium with its surrounding thermostat. The spontaneously 
scattered quantum light features, in this process, were addressed for the case when the dipole-dipole interaction strength among 
the two-level radiators is commensurable, but still smaller, than the transition frequency of a single emitter, respectively. We have
obtained the corresponding master equation describing this multi-qubit ensemble and solved it analiticaly in the steady-state 
which characterizes the established thermal equilibrium. Furthermore, we have demonstrated that the thermal driven small 
two-level ensemble emits light which has sub-Poissonian statistics for weaker thermal baths. For negligible dipole-dipole couplings, 
it is known that this photon statistics occurs for few emitters only and stronger thermal baths that may prevent its detection 
because of the thermal photon background. The collectively spontaneously scattered photon intensity enhances for stronger 
as well as weaker thermal baths due to dipole-dipole interactions among the involved two-level emitters, while suppresses for 
moderately intense to moderately weak intensities of the environmental heat reservoirs, respectively.

This paper is organized as follows. In Sec.~\ref{theo} we describe the analytical approach and the system of interest, while in 
Sec.~\ref{PhCr} we discuss the first- and the second-order photon correlation functions, respectively. Sec.~\ref{RD} presents 
and analyses the obtained results. The article concludes with a summary given in Sec.~\ref{sum}.

\section{Theoretical framework \label{theo}}
We consider an ensemble of $N$ dipole-dipole coupled two-level emitters, each having the transition frequency $\omega_{0}$, 
and interacting with the environmental thermal reservoir at temperature $T$. The atomic subsystem is being densely packed 
so that its linear dimensions are smaller than the photon emission wavelength $\lambda$, i.e.  within the Dicke approximation 
\cite{dicke,gsag,supm1}. The static dipole-dipole coupling strength $\delta \sim d^{2}/r^{3}$ \cite{qab}, where $r$ is the 
mean distance among any of atomic pairs characterized by the dipole $d$, is strong such that the ratio $\delta/\omega_{0}<1$ 
is not negligible and may play a relevant role over the established thermal equilibrium features. Hence, the Hamiltonian describing 
this system in the dipole and rotating-wave approximations \cite{dicke,gsag,supm1,supm2,puri} can be represented as follows: 
$H=H_{0} + H_{i}$, where
\begin{eqnarray}
H_{0}=\sum_{k}\hbar\omega_{k}a^{\dagger}_{k}a_{k} + \hbar\bar \omega_{0}S_{z}-\hbar\tilde\delta S^{+}S^{-},
\label{h0} 
\end{eqnarray}
with $\bar\omega_{0}= \omega_{0} + \tilde \delta$ and $\tilde \delta = \delta/(N-1)$, while
\begin{eqnarray}
H_{i} =i\sum_{k}\bigl\{(\vec g_{k}\cdot \vec d)a^{\dagger}_{k}S^{-} - H.c.\bigr\}. \label{hi}
\end{eqnarray}
The free energies of the environmental electromagnetic field (EMF) thermal modes and atomic subsystems are represented by the 
first two terms of the Hamiltonian $(\ref{h0})$, respectively. The dipole-dipole interaction Hamiltonian is given by the third term 
of this Hamiltonian, i.e. $H_{0}$. Note here that the Hamiltonian describing many two-level radiators is an additive function, i.e. it 
consists from a sum of individual Hamiltonians, characterizing separately each two-level emitters. From this reason, the dipole-dipole 
coupling strength $\delta$ was divided on $N-1$, since there are $N(N-1)$ terms describing the dipole-dipole interacting atoms. 
The Hamiltonian $H_{i}$ accounts for the interaction of the whole ensemble with the environmental thermostat. There, $\vec g_{k}$
=$\sqrt{2\pi\hbar\omega_{k}/V}\vec e_{p}$ is the coupling strength among the few-level emitters and the thermal EMF modes. 
Here, $\vec e_{p}$ is the photon polarization vector with $p \in \{1,2\}$ and $V$ is the quantization volume, respectively. 

In the above Hamiltonians, i.e. (\ref{h0},\ref{hi}), the collective atomic operators $S^{+} = \sum^{N}_{j=1}|e\rangle_{j}{}_{j}\langle g|$ 
and $S^{-}=[S^{+}]^{\dagger}$ obey the usual commutation relations for su(2) algebra, namely, $[S^{+},S^{-}] =2S_{z}$ and 
$[S_{z},S^{\pm}]=\pm S^{\pm}$, where $S_{z} = \sum^{N}_{j=1}(|e\rangle_{j}{}_{j}\langle e| - |g\rangle_{j}{}_{j}\langle g|)/2$ 
is the bare-state inversion operator. Here, $|e\rangle_{j}$ and $|g\rangle_{j}$ are the excited and ground state of the emitter $j$, 
respectively, while $a^{\dagger}_{k}$ and $a_{k}$ are the creation and the annihilation operators of the environmental EMF thermal 
reservoir which satisfy the standard bosonic commutation relations, that is, $[a_{k}, a^{\dagger}_{k'}] = \delta_{kk'}$, and 
$[a_{k},a_{k'}]$=$[a^{\dagger}_{k},a^{\dagger}_{k'}] = 0$ \cite{wm}.

The general form of the master equation, describing the atomic subsystem alone in the interaction picture, is given by
\cite{puri}
\begin{eqnarray}
\frac{d}{dt}\rho(t) =- \frac{1}{\hbar^{2}}{\rm Tr_{f}}\biggl\{\int^{t}_{0}dt'\bigl[H_{I}(t),[H_{I}(t'),\rho(t')]\bigr]\biggr\}, \label{mq}
\end{eqnarray}
where 
\begin{eqnarray}
H_{I}(t) = U(t)H_{i}U^{-1}(t),
\end{eqnarray}
with $U(t)=e^{iH_{0}t/\hbar}$, and the notation ${\rm Tr_{f}}\{ \cdots \}$ means the trace over the thermal EMF degrees of freedom. 
The explicit expression of the interaction Hamiltonian, in the interaction picture is then
\begin{eqnarray}
H_{I} = i\sum_{k}(\vec g_{k}\cdot\vec d)a^{\dagger}_{k}e^{i(\omega_{k}-\hat\omega)t}S^{-} + H.c.,
\label{Hit}
\end{eqnarray}
where 
\begin{eqnarray}
\hat\omega=\bar\omega_{0} + 2\tilde\delta S_{z}. \label{wb}
\end{eqnarray}
Next, one substitutes the Hamiltonian (\ref{Hit}) in the equation (\ref{mq}) and performs the trace over the thermal EMF degrees 
of freedom taking as well into account the operator nature of the exponential function. Subsequently, under the Born-Markoff 
approximations, one arrives at the following master equation describing the atomic subsystem only
\begin{eqnarray}
&{}&\frac{d}{dt}\rho(t) = \nonumber \\
&-& \bigl[S^{+},\frac{\hat \Gamma(\hat \omega)}{2}\bigl(1+\bar n(\hat\omega)\bigr)S^{-}\rho \bigr] - 
\bigl[S^{-},S^{+}\frac{\hat \Gamma(\hat \omega)}{2}\bar n(\hat\omega)\rho \bigr] \nonumber \\
&+& H.c.. \label{fMeq}
\end{eqnarray}
Here $\hat \Gamma(\hat \omega)=2d^{2}\hat\omega^{3}/(3\hbar c^{3})$ 
is the emitters spontaneous decay rate while $\bar n(\hat\omega)=[\exp{(\hbar\hat\omega/k_{B}T)}-1]^{-1}$ is the mean thermal 
photon number, both at the eigenvalues of $\hat\omega$, respectively. For a negligible ratio of the dipole-dipole coupling strength 
over the transition frequency $\omega_{0}$, i.e. $\delta/\omega_{0} \to 0$, the master equation (\ref{fMeq}) turns into the usual 
master equation describing a collection of two-level atoms collectively interacting via their thermal reservoir \cite{gsag,puri,hassan} in 
the Dicke limit, namely, $\dot \rho(t)$=$-\Gamma(\omega_{0})\bigl(1+ \bar n(\omega_{0})\bigr)[S^{+},S^{-}\rho]/2 -
\Gamma(\omega_{0})\bar n(\omega_{0})[S^{-},S^{+}\rho]/2 + H.c.$, where the overdot means differentiation with respect to time. 

While finding the general solution of the master equation (\ref{fMeq}) is not a trivial task at all for a many-atom sample, it possesses a 
steady-state solution. One can observe, by a direct substitution in the Eq.~(\ref{fMeq}), that the expression
\begin{eqnarray}
\rho_{s} = Z^{-1}e^{-\beta H_{a}}, \label{sss}
\end{eqnarray}
is a steady-state solution of it with the parameter $Z$ being the normalization constant determined by the requirement 
${\rm Tr\{\rho_{s}\}=1}$. Here, 
\begin{eqnarray}
H_{a}=\hbar\bar\omega_{0}S_{z} - \hbar\tilde\delta S^{+}S^{-}, \label{ha}
\end{eqnarray}
and $\beta=(k_{B}T)^{-1}$ whereas $k_{B}$ is the Boltzmann's constant. Considering an atomic coherent state $|n\rangle$, denoting 
a symmetrized N-atom state in which $N-n$ particles are in the lower state $|g\rangle$ and $n$ atoms are excited to their upper state 
$|e\rangle$, and that $S^{-}|n\rangle=\sqrt{n(N-n+1)}|n-1\rangle$, $S^{+}|n\rangle = \sqrt{(N-n)(n+1)}|n+1\rangle$ and 
$S_{z}|n\rangle=(n-N/2)|n\rangle$ \cite{puri,hassan}, one can calculate the steady-state expectation values of any atomic correlators 
of interest. 

Therefore, in the next Sections, we shall focus on the quantum statistics of the spontaneously scattered photons in the process of 
dipole-dipole interacting two-level emitters via their environmental thermal reservoir. In this regard, we introduce, respectively, the 
first- and the second-order photon correlation functions describing this feature.

\section{Photon correlation functions \label{PhCr}}
The first- and the unnormalized second-order photon correlation functions at position $\vec R$ can be represented as follows 
\cite{wm,gb1,gb2,LD}
\begin{eqnarray}
G_{1}(\vec R,t) &=& \langle \vec E^{(-)}(\vec R,t)\vec E^{(+)}(\vec R,t) \rangle, \nonumber \\
G_{2}(\vec R,t) &=& \langle: \vec E^{(-)}(\vec R,t)\vec E^{(+)}(\vec R,t) \vec E^{(-)}(\vec R,t)\vec E^{(+)}(\vec R,t):\rangle, 
\nonumber \\ \label{kr}
\end{eqnarray}
where 
\begin{eqnarray}
\vec E^{(-)}(\vec R,t) &=& \sum_{k}\vec g_{k}a^{\dagger}_{k}(t)e^{-i \vec k\cdot\vec R}, \nonumber \\
\vec E^{(+)}(\vec R,t) &=& \sum_{k}\vec g_{k}a_{k}(t)e^{i \vec k\cdot\vec R}, \label{ekr}
\end{eqnarray}
while $:f(\vec R,t):$ indicates normal ordering. In the far-zone limit of experimental interest $R=|\vec R| \gg \lambda$, one can express the first- 
and second-order correlation functions via the collective atomic operators. In this respect, one solves formally the Heisenberg equations for EMF 
operators $\{a^{\dagger}_{k}(t),a_{k}(t)$ using the Hamiltonian (\ref{Hit}). Then, introducing those solutions in the Exps.~(\ref{kr}) and 
integrating over all involved variables, see e.g. \cite{nkmm}, one arrives finally at
\begin{eqnarray}
G_{1}(\vec R,t)&=&\Psi(R)\langle S^{+}(t)\hat \omega^{4}S^{-}(t)\rangle, \nonumber \\
G_{2}(\vec R,t)&=&\Psi(R)^{2}\langle S^{+}(t)\hat \omega^{2}S^{+}(t)\hat \omega^{4}S^{-}(t)\hat \omega^{2}S^{-}(t)\rangle. 
\nonumber \\ \label{krf}
\end{eqnarray}
Here $\Psi(R)=d^{2}(1-\cos^{2}{\xi})/(c^{4}R^{2})$, while $\xi$ is the angle between the direction of vector $\vec R$ and the dipole 
$\vec d$. 
\begin{figure}[t]
\includegraphics[width =7cm]{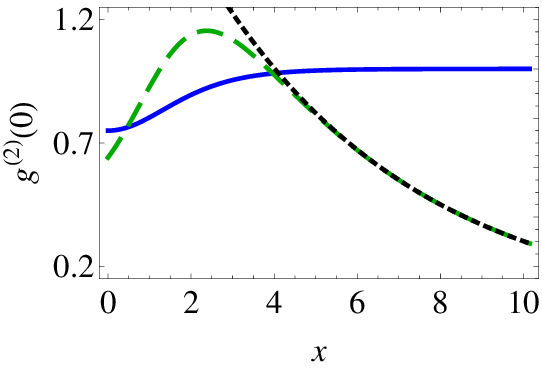}
\caption{\label{fig-1}
The steady-state second-order photon-photon correlation function $g^{(2)}(0)$ as a function of $x=\hbar\omega_{0}/(k_{B}T)$ for a 
two-atom system, $N=2$. Here the solid lines is plotted for a negligible dipole-dipole coupling strength, i.e. $\delta/\omega_{0} \to 0$, 
whereas the dashed one stands for $\delta/\omega_{0} = 0.1$. The short-dashed curve is plotted based on Exp.~(\ref{rsn2}), which 
is valid for $x \gg 1$.}
\end{figure}

The normalized second-order photon-photon correlation function, defined in the usual way, namely,
\begin{eqnarray}
g^{(2)}(t) = \frac{G_{2}(\vec R,t)}{G_{1}(\vec R,t)^{2}},
\label{tkrf}
\end{eqnarray}
takes the next form in the steady-state,
\begin{eqnarray}
g^{(2)}(0) = \frac{\langle S^{+}\hat \omega^{2}S^{+}\hat \omega^{4}S^{-}\hat \omega^{2}S^{-}\rangle}
{\langle S^{+}\hat \omega^{4}S^{-}\rangle^{2}}.
\label{2krf}
\end{eqnarray}
Note that, $g^{(2)}(0)<1$ characterizes sub-Poissonian,  $g^{(2)}(0)>1$ super-Poissonian, and $g^{(2)}(0)=1$ Poissonian photon 
statistics.

In the following Section, we shall investigate the second-order photon-photon correlation function (\ref{2krf}) with the help of the 
steady-state solution (\ref{sss}). Actually, the focus will be on comparing $g^{(2)}(0)$ when $\delta/\omega_{0} \to 0$, a case 
which is known in the scientific literature, with the situation considered here, namely, when $\delta/\omega_{0}\not=0$ but still 
$\delta/\omega_{0} <1$.
\begin{figure}[t]
\includegraphics[width =7cm]{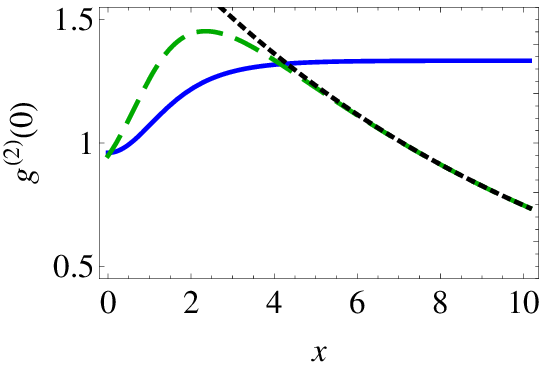}
\caption{\label{fig-2}
Same as in Fig.~(\ref{fig-1}), but for $N=3$.}
\end{figure}

\section{Results and Discussion \label{RD}}
We start our discussions by presenting some limiting cases of the second order photon-photon correlation function given by Exp.~(\ref{2krf}).
Particularly, for $\eta \to 0$, where 
\begin{eqnarray*}
\eta = \frac{\delta}{\omega_{0}}, 
\end{eqnarray*}
we recover the known results (see e.g. \cite{hassan}), namely,
\begin{eqnarray}
g^{(2)}(0)&=&\frac{6(N+3)(N-1)}{5N(N+2)}, ~~{\rm if} ~x \to 0, \nonumber \\
g^{(2)}(0)&=& 2 - \frac{2}{N}, ~~{\rm when} ~x \gg 1,
\label{rsv}
\end{eqnarray}
where
\begin{eqnarray*}
x=\frac{\hbar\omega_{0}}{k_{B}T}. 
\end{eqnarray*}
However, for $\eta \not=0$ and strong thermal baths, i.e. $x \to 0$, and to the second order in the small parameter $\eta$, 
we have obtained that
\begin{eqnarray}
g^{(2)}(0)&=&\frac{6(N+3)(N-1)}{5N(N+2)} \nonumber \\
&+& \frac{48(N+3)(N^{2}+2N-18)\eta^{2}}{25N(N-1)(N+2)}.
\label{rsn1}
\end{eqnarray}
Contrary, for weaker thermal baths when $x \gg 1$, we have obtained the following expression for the second-order photon-photon 
correlation function
\begin{eqnarray}
g^{(2)}(0)&=&2\biggl(1-\frac{1}{N}\biggr)\biggl(\frac{N-1-(N-3)\eta}{(N-1)(1-\eta)}\biggr)^{4} \nonumber \\
&\times&\exp{\biggl(-\frac{2\eta x}{N-1}\biggr)}. \label{rsn2}
\end{eqnarray}
Notice that Exp.~(\ref{rsn1}) and (\ref{rsn2}) turn, respectively,  into those given by Exp.~(\ref{rsv}), when $\eta \to 0$. Moreover, 
the photon statistics, characterized by Exp.~(\ref{rsn2}) for larger values of the parameter $x$, differs considerably between the cases 
$\eta=0$ and $\eta \not=0$, respectively.
\begin{figure}[t]
\includegraphics[width =7cm]{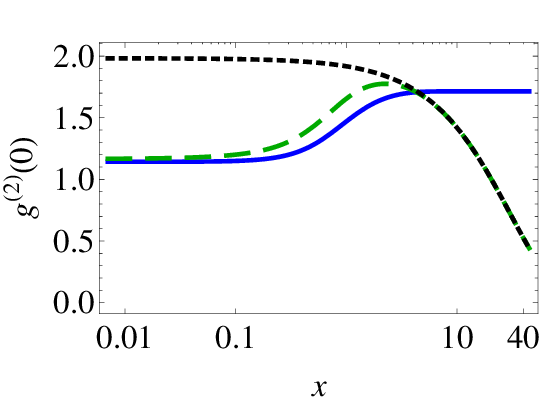}
\caption{\label{fig-3}
The log-linear plot of the steady-state second-order photon-photon correlation function $g^{(2)}(0)$ as a function of
$x = \hbar\omega_{0}/(k_{B}T)$, for N = 7. Other involved parameters are as in Fig.~({\ref{fig-1}}).}
\end{figure}

Figures (\ref{fig-1}) and (\ref{fig-2}) depict the photon statistics of the spontaneously scattered photons by a dipole-dipole interacting 
few-atom system, i.e. for $N=2$ and $N=3$, respectively, via their environmental thermostat. Particularly, the solid lines are plotted for 
a negligible dipole-dipole coupling strength, i.e. $\delta/\omega_{0} \to 0$, whereas the dashed ones stand for $\delta/\omega_{0}=0.1$.
The short-dashed curves represent the Exp.~(\ref{rsn2}), which is valid for $x \gg 1$. Interestingly, the photon statistics turns into 
a quantum photon statistics, i.e. $g^{(2)}(0)<1$, when $\eta \not =0$ and for weaker thermal baths, $x \gg 1$. This behaviour is
typical for several dipole-dipole interacting atoms, too. From this reason, in Fig.~(\ref{fig-3}) we plot the steady-state second-order 
photon-photon correlation function $g^{(2)}(0)$ for $N = 7$ dipole-dipole interacting atoms which supports it. Thus the thermal 
mediated dipole-dipole interactions among two-level emitters in a several-atom Dicke-like sample are responsible for the sub-Poissonian 
photon statistics of the spontaneously scattered photons. In the absence of the dipole-dipole interactions, that is when $\eta \to 0$, 
sub-Poissonian photon statistics occurs in this system only for $N \in \{1,2,3\}$ and for strong thermal baths, $x \ll 1$, see 
Fig.~(\ref{fig-1}) and Fig.~(\ref{fig-2}). Also, one can observe a nice concordance with the limiting cases of the second-order 
photon-photon correlation function $g^{(2)}(0)$, given by the Exps.~(\ref{rsv}-\ref{rsn2}), and the plots in Figs.~(\ref{fig-1}-\ref{fig-3}).
\begin{figure}[b]
\includegraphics[width =7cm]{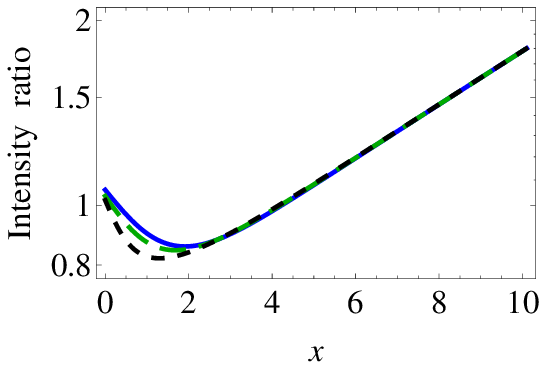}
\caption{\label{fig-4}
The log-plot of the ratio $G_{1}(\vec R,\eta \not=0)/G_{1}(\vec R,\eta=0)$ as a function of $x = \hbar\omega_{0}/(k_{B}T)$ 
for $\eta=0.1$. The solid line is plotted for $N=2$, the dashed one for $N=3$, while the short-dashed curve for $N=7$, 
respectively.}
\end{figure}

While this tendency, i.e. occurring of sub-Poissonian photon statistics for bigger values of $x=\hbar\omega_{0}/(k_{B}T)$, persists 
for larger atomic ensembles, it will be less useful since the photon flux might be too weak in this case. To clarify this issue, one 
represents the cooperative intensity of spontaneously scattered EMF by the dipole-dipole interacting two-level emitters in a 
weaker thermal environment, i.e. for $x \gg 1$, as follows
\begin{eqnarray}
G_{1}(\vec R)/\Psi(R)= N\bigl(1-\eta\bigr)^{4}\exp{\bigl(-x(1-\eta)\bigr)}. \label{G1a}
\end{eqnarray}
On the other hand, from the Exp.~(\ref{rsn2}) when setting $g^{(2)}(0)=1$, one can obtain the $\eta'{\rm s}$ ranges in order 
the sub-Poissonian photon statistics to occur for $N \gg1$ and $x \gg 1$, namely,
\begin{eqnarray}
\eta > (N/x)\ln{(\sqrt{2})}. \label{etc}
\end{eqnarray}
Note here that $\eta = (N/x)\ln{(\sqrt{2})}$ is the condition for Poissonian, whereas $\eta < (N/x)\ln{(\sqrt{2})}$ 
for super-Poissonian photon statistics, respectively. Although the Exp.~(\ref{etc}) is not difficult to fulfil, the photon 
intensity, given by Exp.~(\ref{G1a}), will be too low for larger ensembles and weaker thermal baths, $x \gg 1$. 
The reason consists in the Bose-Einstein nature of the atomic statistics, meaning that in a thermal environment at 
equilibrium,  the two-level emitters tend to reside in their ground state for $\{N,x\} \gg 1$. Under these circumstances, 
less atoms get excited and, therefore, the spontaneously generated EMF might be weak, as well. Finally, the photon 
statistics changes from sub-Poissonian to super-Poissonian if $\delta \to -\delta$ and $x \gg 1$.

Note that, from Exp.~(\ref{G1a}) follows that for weaker thermal baths when $x \gg 1$, $G_{1}(\vec R,\eta \not=0) 
> G_{1}(\vec R,\eta=0)$, see Fig.~(\ref{fig-4}). However, from the general expression of the intensity, i.e. 
Exp.~(\ref{krf}), one has that $G_{1}(\vec R,\eta \not=0) < G_{1}(\vec R,\eta=0)$, within moderate weak, 
$x > 1$, to moderate intense, $x \le 1$ heat baths, respectively. This will turn again to 
$G_{1}(\vec R,\eta \not=0) > G_{1}(\vec R,\eta=0)$ for stronger thermal baths, i.e. when $x\to 0$, see 
Fig.~(\ref{fig-4}). Particularly, for stronger environmental thermal baths intensities when $x \to 0$, one has that
\begin{eqnarray}
\frac{G_{1}(\vec R,\eta \not= 0)}{G_{1}(\vec R,\eta = 0)}=1+ 72\eta^{2} + 2064\eta^{4}/7. \label{rtit}
\end{eqnarray}

Generalizing, the thermal mediated dipole-dipole interactions are responsible for sub-Poissonian statistics of the 
spontaneously scattered photons by a small two-level ensemble, in a weak thermal bath and within the Dicke 
limit. Furthermore, due to presence of the dipole-dipole interactions between the two-level emitters, the 
collectively scattered spontaneous emission light suppresses or enhances depending on the intensities of the 
environmental thermal baths, see also \cite{phsupr}. In principle, the relationship among the quantum nature 
of the spontaneously scattered photons in this process and the ensemble's quantum thermodynamics would 
be an interesting issue to be addressed in future investigations. 

\section{Summary \label{sum}}
We have investigated the steady-state quantum dynamics of an ensemble of dipole-dipole interacting two-level emitters, 
in the Dicke limit, which is mediated by the environmental thermostat. Particularly, the focus was on the quantum nature 
of the collective spontaneously scattered photons in this process. We have demonstrated that the dipole-dipole interaction 
among the two-level emitters is responsible for sub-Poissonian photon statistics scattered by a small ensemble consisting 
from few to several atoms, respectively. The quantum light effect occurs when the energy of a free atom is larger than 
the corresponding one due to the thermal reservoir. This quantum future survives for larger ensembles, however, the 
photon intensity might be too weak in this case. Without the dipole-dipole interactions among the two-level radiators, 
the sub-Poissonian statistics occurs for few atoms and stronger thermal baths only. The collectively scattered spontaneous 
emission light intensity suppresses or enhances depending on the intensities of the environmental thermal baths.

\acknowledgments
The financial support from the Moldavian Ministry of Education and Research, via grant No. 011205, is gratefully acknowledged.


\end{document}